\documentclass[aps,12pt,nofootinbib,showpacs,amssymb]{revtex4}

\usepackage{graphicx}
\usepackage{amsmath,amsfonts}

\DeclareMathOperator{\Tr}{Tr}

\newcommand{\nn}{\nonumber}

\newcommand{\be}{\begin{equation}}
\newcommand{\ee}{\end{equation}}
\newcommand\ba{\begin{eqnarray}}
\newcommand\ea{\end{eqnarray}}

\begin{document}
\title{Bremsstrahlung photon polarization for $ee^\pm\to (e\gamma)e^\pm$, and
$ep\to (e\gamma)p$ high energy collisions}

\author{S.~Bakmaev}
\affiliation{Joint Institute of Nuclear Research, 141980 Dubna, Russia}

\author{Yu.~M.~Bystritskiy}
\affiliation{Joint Institute of Nuclear Research, 141980 Dubna, Russia}

\author{E.~A.~Kuraev}
\affiliation{Joint Institute of Nuclear Research, 141980 Dubna, Russia}

\author{E.~Tomasi-Gustafsson}
\affiliation{DAPNIA/SPhN,CEA/Saclay,91191 Gif-sur Yvette, Cedex, France}

\date{\today}

\begin{abstract}

The  polarization of bremsstrahlung photon in the processes $ee^\pm\to (e\gamma)e^\pm$,
 and $ep\to (e\gamma)p$ is calculated for peripheral kinematics, in the high energy
limit where the cross section does not decrease with the incident energy.
When the initial electron is unpolarized(longitudinally polarized) the final photon can
be linearly (circularly) polarized. The Stokes parameters of the photon polarization
are calculated as a function of the kinematical variables of process: the energy of
recoil particle, the energy fraction of scattered electron, and the polar and azimuthal
angles of photon. Numerical results are  given in form of tables, for typical values
of the relevant kinematic variables.

\end{abstract}

\maketitle

\section{Introduction}
It is known that the bremsstrahlung photon in electron-positron and electron-proton
scattering can be polarized \cite{Fano}.
If the initial electron is unpolarized the bremsstrahlung photon may acquire linear
polarization; if the initial electron is longitudinally polarized then the photon may
 acquire left or right circular polarization.

Polarized beams allow to access more detailed information
about the target properties in comparison with unpolarized reactions,
which give differential cross sections averaged over the amplitudes. As an example,
the circular polarization of bremsstrahlung photon in the scattering of charged leptons
contains information about the standard model concerning heavy vector bosons \cite{Po83}:
one could detect neutral current effects by looking to the helicity of the outgoing
particles emitted in unpolarized fermion scattering.

To measure the degree of polarizationof the photons, specific polarimeters should be used,
based on physical processes
with sufficiently large cross sections and analyzing power.

Linearly polarized photons produce bremsstrahlung photons which appear in
scattering of electrons on  a crystal surface (see \cite{Boldyshev},\cite{Kondo}).
The degree of photon polarization obtained in such a way depends on many
external parameters, including the crystal characteristics. Photoproduction of
electron pairs in triplet state has been suggested as a possible way to measure
the degree of polarization of photon beams, in a wide energy range  \cite{Boldyshev}
and it was shown that the analyzing powers may reach 14\% \cite{Boldyshev} .

In the present work, we calculate the polarization of bremsstrahlung photons from
electron scattering on electron or proton in the high energy limit. The cross section
and the degree of polarization, being sufficiently large, we suggest that this process
could be used for electron polarimetry.

Let us  consider the bremsstrahlung process for the scattering of an electron on an
electron(positron) or on a proton:
\ba
e^-(p_1,\lambda)+P(p)\to e^-(p'_1)+\gamma(k,e)+P(p'),
\label{eq:eq1}
\ea
where $\lambda$ is the longitudinal polarization and $p_1$ the momentum of the incident
electron, with $p_1^2=m^2$ ($m$ is the mass of electron), $k$ is the momentum of the
photon ($k^2=0$), and $e$ its polarization vector. $P$ denotes the target particle, $p^2=p'^2=M^2$, where $M=M_p$ ($m$) for the case of a proton ($e^\pm$) target.

We will consider the kinematics related to peripheral collisions, where particles
of energy $E$ scatter on a target of mass $M$ at small angles, in the laboratory system.
This kinematical regime is characterized by
$$s=(p+p_1)^2-M^2=2ME \gg M^2\sim |q^2|,~ q=p-p'.$$
In peripheral kinematic and in the high energy limit, the cross sections of particle
production do not depend on the incident energy. It is convenient to use the  Sudakov's
parametrization for the kinematical variables, as defined below. The laboratory frame is
taken as the reference frame all along the
paper.

\section{Formalism}
\subsection{Kinematics}
\label{AppendixKinematics}
Considering  peripheral processes, it is
convenient to use the Sudakov's parametrization of momenta. Any four-vector, $v$, can be
represented as $v=(v_0,v_{\parallel},\vec v_{\perp})$, where $v_0$ is the time component,
 $v_{\parallel}$ is the longitudinal component with respect to the momentum of the
initial electron, and $v_{\bot}$ is the two-dimensional vector of the transversal
component. Let us introduce two light-like four-vectors,
$\tilde p=p-p_1M^2/s$, $\tilde p_1 = p_1-pm^2/s\simeq p_1$, with $s=2pp_1$.
The explicit components of these vectors are $p_1=E(1,1,0,0)$
and $\tilde p= (M/2)(1,-1,0,0)$.

Any vector can be expressed in a basis defined by
$\tilde p$, and $p_1$, with the help of the coefficients $\alpha_i$ and  $\beta_i$:
\begin{subequations} \label{E:vp}
\begin{gather}
k=\bar x p_1+\alpha_{\gamma}\tilde{p}+k_\bot, \label{E:vp1} \\
p'_1=xp_1+\alpha_e\tilde{p}+p_{\perp}, \label{E:vp2} \\
q=\beta_q p_1+\alpha_q\tilde{p}+q_{\perp}, \label{E:vp3} \\
e(k)=\beta p_1+\alpha\tilde{p}+e_\bot. \label{E:vp4}
\end{gather}
\end{subequations}
where $x$ is the  fraction of initial energy carried by the
scattered electron, $\bar x = 1-x$ is the energy fraction carried by the photon and
$p_{\perp}^2=-\vec p^{\,\,2}$, $k_{\perp}^2=-\vec k^{\,\,2}$,
$q_{\perp}^2=-\vec q^{\,\,2}$, $e_{\perp}^2=-\vec e^{\,\,2}$,
$\vec q = \vec p + \vec k$ correspond to the components of the vectors
$p'_1$, $k$, $q$, $e$ which are orthogonal to the vectors
$\tilde p$ and $\tilde p_1$. Here $\vec p$, $\vec k$, $\vec q$, $\vec e$ are
two-dimensional vectors.

Applying on-mass shell conditions and gauge invariance: $e(k) k = 0$, one finds
the following relations:
\begin{subequations} \label{E:gp}
\begin{gather}
    d_1 = 2 p_1 k = \frac{1}{\bar{x}} \left(m^2 \bar{x}^2 + {\vec k}^{\,2}\right)
    = \frac{\Delta_1}{\bar{x}},
    \qquad
    2 p_1 e = \frac{2}{\bar{x}} \vec k \vec e,  \label{E:gp1} \\
    d_2 = 2 p_1' k =\frac{1}{x \bar{x}} \left(m^2 \bar{x}^2 + {\vec b}^{\,2}\right)
    = \frac{\Delta_2}{x \bar{x}},
    \qquad
    2 p_1' e = \frac{2}{\bar{x}} \vec b \vec e, \label{E:gp2} \\
    2 p p_1' = \frac{1}{x} \left(m^2 (1+x^2) + {\vec p}^{\,2}\right),
    \qquad \vec b = x \vec q - \vec p.  \label{E:gp3}
\end{gather}
\end{subequations}
where we introduced the termes
${\Delta_1}=\left(m^2 \bar{x}^2 + {\vec k}^{\,2}\right)$ and ${\Delta_2}= \left(m^2 \bar{x}^2 + {\vec b}^{\,2}\right)$.
\begin{figure}[t]
\begin{center}
\includegraphics[bb=0 0 470 555, scale=0.22]{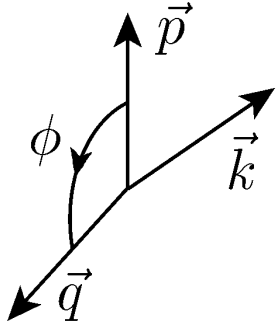}
\caption{Definition of azimuthal angle $\phi$ of recoil particle.}
\label{FigAnglePhi}
\end{center}
\end{figure}
The phase volume of the final state
\be
    d\Gamma = (2\pi)^{-5}
    \frac{d^3 k}{2\omega}\frac{d^3 p_1'}{2E_1'}\frac{d^3 p'}{2 E'}
    \delta^4\left( p+p_1 - p_1'-p'-k \right), \label{eq:eqps}
\ee
after introducing an auxiliary integration
\be
\delta^4\left( p+p_1 - p_1'-p'-k \right) =
d^4q \delta^4\left( p_1+q - k - p_1' \right) \delta^4\left( q - p + p' \right),
\ee
can be expressed in terms of Sudakov's variables as :
\ba
d\Gamma=\frac{d^2q \,d^2p \, dx}{4sx\bar x(2\pi)^5}
=\frac{E'dE'dx\theta_ed\theta_e d\phi}{2^7\bar{x}\pi^4},
\label{eq:eqpg}
\ea
where $\phi=(\widehat{\vec{p},\vec{q}})$ is the azimuthal angle
(see Fig. \ref{FigAnglePhi}), delimited by two planes:
the plane which contain the momenta of the initial and final electrons,
($p_1$,$p_1'$) and the plane defined by the momenta of the initial electron and
of the exchanged photon ($p_1$,$q$),
$\theta_e=|\vec{p}|/(Ex)$ is the angle between initial and scattered
electron directions.

In the Laboratory frame
the transverse momentum $|\vec{q}|$ is related to the energy of the recoil particle.
Due to four-momentum conservation the recoil proton momentum  $p'=(E',\vec{p}{\,'})$
can be written as:
\be
\vec{p}{\,'}^2=\vec{q}^{\,2}+\frac{(\vec{q}^{\,2})^2}{4M^2},
\ee
and the following relations hold:
\be
E'^{\,2}=\frac{(\vec{q}^{\,2}+2M^2_p)^2}{4M^2_p}, ~
\vec{p}^{\,\,2}=(E\theta_e x)^2,~ \vec{q}^{\,\,2}=2M^2_p+2M_pE'.
\ee
\subsection{Unpolarized electron}

According to the formalism developed in Ref. \cite{BFKK81}, the matrix element
of the process (\ref{eq:eq1}) (which is illustrated in Fig. \ref{FigFD}), for not
negligible momentum transferred by the exchanged virtual photon, $|\vec{q}|$, can be
written as:
\begin{figure}[t]
\begin{center}
\includegraphics[bb=0 0 1992 523, scale=0.22]{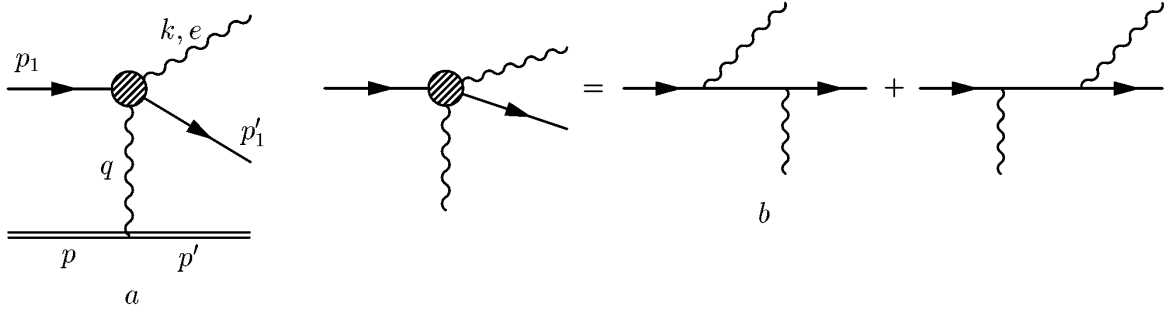}
\caption{Feynman diagrams which are relevant in
peripheral kinematics.}
\label{FigFD}
\end{center}
\end{figure}
\be
{\cal M}=\frac{(4\pi\alpha)^\frac{3}{2}}{\vec q^{\,\,2}}\frac{2}{s}
\left[\bar u(p')\Gamma_\lambda p^\lambda_1u(p)\right]
\left[\bar{u}(p'_1)O_{\mu\nu}\tilde{p}^\mu e^\nu u(p_1)\right],
\ee
where $\Gamma_\mu = \gamma_\mu$ in case of $e^\pm$ target, and
$\Gamma_\mu = F_1(q^2)\gamma_\mu + (1/4M_p)(\hat q \gamma_\mu- \gamma_\mu\hat q) F_2(q^2)$
for a proton target, where
$F_{1,2}(q^2)$ are the proton form factors. The vectors $\vec k$, $\vec q$ and $\tilde p$ are
defined in Eqs. (\ref{E:vp}).

After summing over the lepton quantum numbers, the square of the matrix element is:
\ba
\sum |{\cal M}|^2 = \frac{\left(\pi \alpha\right)^3}{\left(\vec q^{\,\,2}\right)^2}
2^{11} s^2 D \frac{1}{4}
\Tr \left[\hat{p}'_1\left(\hat{e}\rho+\frac{\hat{e}\hat{q}\hat{\tilde{p}}}{sd_2}
-\frac{\hat{\tilde{p}}\hat{q}\hat{e}}{sd_1}\right)\hat{p}_1\left(\rho \hat{e^*}+
\frac{\hat{\tilde{p}}\hat{q}\hat{e^*}}{sd_2}-\frac{\hat{e^*}\hat{q}\hat{\tilde{p}}}{sd_1}\right)
\right],
\label{eq:eqm2}
\ea
where $\rho = \frac{1}{d_2} - \frac{x}{d_1}$ and
\be
D=\left\{\begin{array}{ll}
           1, &\mbox{ for~electron target,}\\
     F_1^2(-\vec{q}^2)+\frac{\vec{q}^2}{4M_p^2}F_2^2(-\vec{q}^2)
       &\mbox{ for~proton target.}\end{array}
\right.\nn
\ee
In case of unpolarized final photon, the cross section can be written as:
\ba
M^3_p \frac{2\pi d\sigma}{dxdE'd\vec p^{\,\,2}d\phi}=
\frac{4\alpha^3}{M^2_p} D R, ~
R=\frac{M^6_p A}{(\vec{q}^{\,\,2})^2x \bar x},
\label{R}
\ea
where
\ba
A&=&(x\bar{x})^2\left(\frac{1}{\Delta_1}-\frac{1}{\Delta_2}\right)^2\left[
\frac{\vec{p}^{\,\,2}}{2x}+\frac{(\vec{q}-\vec{p})(x\vec{q}-\vec{p})}{\bar x^2}\right]+\nn\\
&&+\frac{(x\bar{x})^2}{\Delta_2}\left(\frac{1}{\Delta_2}-\frac{1}{\Delta_1}\right)\left[
-\vec{p}\vec{q}-\frac{2}{\bar{x}}(x\vec{q}^{\,\,2}-\vec{p}\vec{q})\right]-\nn\\
&&-\frac{x \bar{x}^2}{\Delta_1}\left(\frac{1}{\Delta_2}-\frac{1}{\Delta_1}\right)
\left[\frac{1}{2}\vec{p}\vec{q}-\frac{2x}{\bar{x}}\vec{q}^{\,\,2}+\frac{2x}{\bar{x}}\vec{p}\vec{q}\right]
+\left[\frac{(x\bar{x})^2}{\Delta^2_2}+\frac{(\bar{x})^2}{\Delta^2_1}\right]\frac{x\vec{q}^{\,\,2}}{2},
\label{A}
\ea
The quantity $R$ is calculated for $x=1/2$ and for different values of $\vec q^2$
and $\vec p^2$ in Table \ref{TableA}.

Introducing the photon polarization density matrix:
\ba
\sum_{ij}e^\lambda_ie^{*\lambda}_j=\frac{1}{2}[I+\xi_1\sigma_1+
\xi_3\sigma_3]_{ij},
\ea
Eq. (\ref{eq:eqm2}) can be written in the form
$$\sum |{\cal M}|^2 \sim A+B\xi_1+C\xi_3=A[1+X_1\xi_1+X_3\xi_3].$$
The polarization state of bremsstrahlung photon is
characterized by the Stokes parameters, which have the form \cite{Akhiezer}
\ba
X_1=\frac{B}{A},\quad X_3=\frac{C}{A}.
\label{X1X3}
\ea
with $B$ and $C$ given by:
\ba
B&=&(x\bar{x})^2\left(\frac{1}{\Delta_2}-\frac{1}{\Delta_1}\right)^2\left[
x\vec{q}^{\,\,2}\sin(2\phi)-(1+x)|\vec{p}||\vec{q}|\sin(\phi)\right]+\nn\\
&&+\frac{(x\bar{x})^2}{\Delta_2}\left(\frac{1}{\Delta_2}-\frac{1}{\Delta_1}\right)
\left[ x\vec{q}^{\,\,2}\sin(2\phi)-|\vec{p}||\vec{q}|\sin(\phi)\right]+\nn\\
&&+\frac{2x^2\bar{x}}{\Delta_1}\left(\frac{1}{\Delta_2}-\frac{1}{\Delta_1}\right)
\left[\vec{q}^{\,\,2}\sin(2\phi)-|\vec{p}||\vec{q}|\sin(\phi)\right],
\ea
\ba
C&=&x^2\left(\frac{1}{\Delta_2}-\frac{1}{\Delta_1}\right)^2
\left[\vec{p}^{\,\,2}+x\vec{q}^{\,\,2}\cos(2\phi)- 2x|\vec{p}||\vec{q}|\cos(\phi)\right]
-\nn\\
&&-\frac{2 x^2 \bar{x}}{\Delta_2}\left(\frac{1}{\Delta_2}-\frac{1}{\Delta_1}\right)
\left[x\vec{q}^{\,\,2}\cos(2\phi)-|\vec{p}||\vec{q}|\cos(\phi)\right]+\nn\\
&&+\frac{2 x^2 \bar{x}}{\Delta_1}\left(\frac{1}{\Delta_2}-\frac{1}{\Delta_1}\right)
\left[\vec{q}^{\,\,2}\cos(2\phi)-|\vec{p}||\vec{q}|\cos(\phi)\right]
-\frac{(x\bar{x})^2}{\Delta_1\Delta_2}\vec{q}^{\,\,2}\cos(2\phi),
\ea
where $\Delta_{1,2}$ are defined in Eqs.(\ref{E:gp1},\ref{E:gp2}).

The quantities $X_1$ and $X_3$ are calculated in
Tables \ref{TableX1} and \ref{TableX3} for typical kinematics. As one can see, the values of $X_1$ are quite constant and very large, of the order of 80\%.
 The values of $X_3$ are also sizeable and increase with energy.
The magnitude of the cross section can be calculated from Eq. (\ref{R}) and Table \ref{TableX1} , the kinematical
coefficient $4\alpha^3/M_p^2$ being of the order of 700~pb.
So the bremsstrahlung with unpolarized initial electron can be used as a
polarimeter one. Really it has large cross section and the
characteristics of linear polarization of photon depends smoothly in
known way on kinematic

\begin{table}[t]
\begin{tabular}{|c|c|c|c|c|c|c|}
\hline
$\vec q^{\,2}$, GeV / $\vec p^{\,2}$, GeV & 0.50 & 1.00 & 1.50 & 2.00 & 2.50 & 3.00\\

\hline
0.50 & 1.583 & 0.489 & 0.225 & 0.127 & 0.080 & 0.055\\
\hline
1.00 & 0.493 & 0.198 & 0.102 & 0.061 & 0.040 & 0.028\\
\hline
1.50 & 0.217 & 0.104 & 0.059 & 0.037 & 0.025 & 0.018\\
\hline
2.00 & 0.114 & 0.062 & 0.037 & 0.025 & 0.017 & 0.013\\
\hline
2.50 & 0.067 & 0.040 & 0.026 & 0.018 & 0.013 & 0.010\\
\hline
3.00 & 0.043 & 0.027 & 0.018 & 0.013 & 0.010 & 0.007\\

\hline
\end{tabular}
\caption{calculated values for the term $R$ (Eq. (\ref{R}))
for $\phi=\pi/2$, $x=1/2$ as a function
of $\vec q^{\,2}$ and $\vec p^{\,2}$.}
\label{TableA}
\end{table}

\begin{table}[t]
\begin{tabular}{|c|c|c|c|c|c|c|}
\hline
$\vec q^{\,2}$, GeV / $\vec p^{\,2}$, GeV & 0.50 & 1.00 & 1.50 & 2.00 & 2.50 & 3.00\\

\hline
0.50 & -0.841 & -0.853 & -0.845 & -0.827 & -0.806 & -0.785\\
\hline
1.00 & -0.816 & -0.841 & -0.851 & -0.853 & -0.851 & -0.845\\
\hline
1.50 & -0.799 & -0.827 & -0.841 & -0.849 & -0.853 & -0.853\\
\hline
2.00 & -0.784 & -0.816 & -0.831 & -0.841 & -0.847 & -0.851\\
\hline
2.50 & -0.769 & -0.807 & -0.823 & -0.833 & -0.841 & -0.846\\
\hline
3.00 & -0.754 & -0.799 & -0.816 & -0.827 & -0.834 & -0.841\\

\hline
\end{tabular}
\caption{The Stokes parameter $X_1$ (Eq. (\ref{X1X3}))
for $\phi=\pi/2$, $x=1/2$ as a function
of $\vec q^{\,2}$ and $\vec p^{\,2}$.}
\label{TableX1}
\end{table}

\begin{table}[t]
\begin{tabular}{|c|c|c|c|c|c|c|}
\hline
$\vec q^{\,2}$, GeV / $\vec p^{\,2}$, GeV & 0.50 & 1.00 & 1.50 & 2.00 & 2.50 & 3.00\\

\hline
0.50 & 0.397 & 0.483 & 0.534 & 0.569 & 0.596 & 0.617\\
\hline
1.00 & 0.308 & 0.397 & 0.447 & 0.483 & 0.511 & 0.534\\
\hline
1.50 & 0.251 & 0.345 & 0.397 & 0.432 & 0.460 & 0.483\\
\hline
2.00 & 0.209 & 0.308 & 0.360 & 0.397 & 0.424 & 0.447\\
\hline
2.50 & 0.175 & 0.277 & 0.332 & 0.369 & 0.397 & 0.419\\
\hline
3.00 & 0.147 & 0.251 & 0.308 & 0.345 & 0.374 & 0.397\\

\hline
\end{tabular}
\caption{The Stokes parameter $X_3$ (Eq. (\ref{X1X3}))
for $\phi=\pi/2$, $x=1/2$ as a function
of $\vec q^{\,2}$ and $\vec p^{\,2}$.}
\label{TableX3}
\end{table}

\subsection{Circular photon polarization}

The longitudinal polarization of the initial electron induces a circular polarization
of the photon \cite{BGGK04}:
\ba
X_2=\frac{|{\cal M}_+|^2-|{\cal M}_-|^2}{|{\cal M}_+|^2+|{\cal M}_-|^2},
\ea
where ${\cal M}_\pm$ is the light-cone projection of matrix element of the subprocess
of creation of a photon with
chirality $\pm$ in the process where the lepton has positive chirality :
$e(p_1,+)+\gamma^* \to \vec e(p'_1,+)+\gamma(k,\pm)$.

\begin{figure}[t]
\begin{center}
\includegraphics[bb=0 0 470 555, scale=0.22]{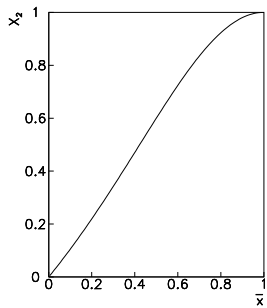}
\caption{The Stokes parameter $X_2$ as a function of the energy
fraction carried by the photon $\bar x$.}
\label{fig:figx2}
\end{center}
\end{figure}

The Stokes parameter $X_2$, related to the circular polarization of the photon
depends only on the energy fraction carried by the photon:
\ba
X_2=\frac{1-x^2}{1+x^2}=\frac{\bar{x}(2-\bar{x})}{2-2\bar{x}+\bar{x}^2}.
\label{eq:eqx2}
\ea
One can see that the larger is the energy of the photon, the larger is the degree of
polarization (Fig. \ref{fig:figx2}).

\section*{Conclusion}
We calculated the linear polarization of the bremsstrahlung photon for unpolarized
high-energy
$e-p$, $e-e^\pm$ scattering at small scattering angles and high energy. The  relevant
Stokes parameters are functions of
kinematical variables such as the transferred momentum, the polar and azymuthal angles
of the  photon and its energy fraction. For the case of
longitudinally polarized initial electrons, the photon has nonzero circular
polarization which depends only on its energy fraction.

The accuracy of formulae given above is determined by radiative
corrections and on the omitted terms. We estimate it as
\ba
    1 + O\left(\frac{\vec p^{\,2}}{s}, \frac{\alpha}{\pi} \ln\frac{\vec p^{\,2}}{m^2}, \frac{m^2}{\vec p^{\,2}}\right).
\ea
As an example, for the upgraded JLab facility ($E \approx 12~\mbox{GeV}$) it is of order
 $\sim~10 \%$.

In this paper we emphasized the possibility to obtain polarized photons in the process of
peripheral scattering of leptons. We derived the expressions that relate the polarization
parameters and the cross section to the relevant kinematical variables and calculated these observables, for different kinematical conditions.

We showed that the cross section and the degree of polarization is sufficiently large and that this process should be taken into consideration for designing photon polarimeters in the GeV range.

It should be stressed that the Stokes parameters that characterize the bremsstrahlung photon polarization do not depend neither on the initial energy nor on the target mass.



\begin{thebibliography}{99}

\bibitem{Fano}
U.Fano Phys. Rev. {\bf 93}, 121 (1954);
  M.~P.~Rekalo and I.~M.~Sitnik,
  Phys.\ Lett.\  B {\bf 356}, 434 (1995).
\bibitem{Po83}
  M.~Porrmann,
  Nucl.\ Phys.\ A {\bf 399} (1983) 365.
%



\bibitem{Boldyshev}
  V.~F.~Boldyshev, E.~A.~Vinokurov, N.~P.~Merenkov and Yu.~P.~Peresunko,
  Phys.\ Part.\ Nucl.\  {\bf 25} (1994) 292 and Refs. therein.

\bibitem{Kondo}
  K.~Kondo {\it et al.},
  Nucl.\ Instrum.\ Meth.\  {\bf 114}, 365 (1974).


\bibitem{BGGK04}
  E.~Bartos, M.~V.~Galynskii, S.~R.~Gevorkyan and E.~A.~Kuraev,
  Nucl.\ Phys.\ B {\bf 676}, 390 (2004).

\bibitem{BFKK81}
  V.~N.~Baier, E.~A.~Kuraev, V.~S.~Fadin and V.~A.~Khoze,
  Phys.\ Rept.\  {\bf 78}, 293 (1981).

\bibitem{Akhiezer} A.I. Akhiezer, V.B. Berestetskii, Quantum Electrodynamics (1981)







\end{thebibliography}
\end{document}